# An Algebraic View on the Semantics of Model Composition

Christoph Herrmann[1], Holger Krahn[1], Bernhard Rumpe[1], Martin Schindler[1], and Steven Völkel[1]

[1] Institute for Software Systems Engineering, Braunschweig University of Technology,
Mühlenpfordtstraße 23
38106 Braunschweig, Germany
{herrmann, krahn, rumpe, m.schindler, voelkel}@sse.cs.tu-bs.de

**Abstract.** Due to the increased complexity of software development projects more and more systems are described by models. The sheer size makes it impractical to describe these systems by a single model. Instead many models are developed that provide several complementary views on the system to be developed. This however leads to a need for compositional models. This paper describes a foundational theory of model composition in form of an algebra to explicitly clarify different variants and uses of composition, their interplay with the semantics of the involved models and their composition operators.

## 1 Model composition

The complexity of software products and therefore of their development projects is steadily increasing. To handle this complexity models are used as an intermediate result to raise the level of abstraction, to enhance the understanding, and to simplify analysis and prediction of properties of the system under development. Nowadays modeling languages like the UML (Unified Modeling Language) and an increasing number of DSLs (Domain Specific Languages) are used for planning, architecting, developing, coding, deploying, and documentation purposes. Based on these languages a number of development approaches like OMG's Model Driven Architecture can be classified as "Model Driven Engineering" (MDE).

In any complex software system, mastering complexity means using a variety of semantically and syntactically precise [1,2] models to describe different aspects and views of the software system. Therefore it is essential to understand how these different models fit together and complement each other. For an integrated understanding, a clear definition of what composition of models means is necessary.

Model composition has impacts on at least three different levels:

- Syntactic level: the way the composition between models can explicitly be expressed as a new model in an appropriate modeling language.
- Semantic level: the meaning of the composed models as a unit in terms of semantics of the modeling languages involved.
- Methodic level: the integration of model composition techniques in software development processes and tools.

A clear explanation of a composition mechanism of models on each of these dimensions is necessary to facilitate a "compositional" use of models in development projects. E.g., for an integrated understanding of some models describing aspects of the same system it is not necessary to provide a syntactic composition operator that explicitly produces an integrated model. Instead it is essential to understand the meaning of "composition" using a semantic composition. For code generation purposes it is however often necessary to explicitly calculate the integrated model, because only from there it is possible to start the generator. This is a pity, because already in 1972 Parnas introduced modularity in his article [3] as an important requisite for independent understanding, development, and compilation – something we have achieved on code level, but not on model level so far. It therefore depends on the form of use which properties a model composition operator must have.

In contrast to concrete model composition techniques [4, 5, 6] we examine in this paper syntactic and specifically semantic properties of model composition as basis for a methodical discussion and therefore regard this paper as a first contribution to a wider discussion on compositionality of models.

The rest of the paper is structured as follows. Section 2 gives a compact recapitulation and introduction to our understanding of syntax and semantics of a modeling language. Section 3 describes the properties of model composition in algebraic terms. We derive requirements for well-defined model composition operators and give a first classification of possible operators. Section 4 describes related work, followed by a conclusion in Section 5.

## 2  Syntax and Semantics of Models

In software engineering we are basically concerned with graphical or textual languages to describe structure, behavior, or interaction of systems, interfaces etc. As these models shall usually be understood by tools, e.g., for code generation and test case definition there must be a clear definition of what the language concepts are. This is in sharp contrast to many other forms of models, where there is no formal and explicit definition of the modeling language used (see, e.g., architectural or medical models).

Formally, a modeling language $M$ is a set of well-formed models. So a model $m \in M$ is syntactically well-formed, both by context-free syntax as well as conforming to all context-conditions. Each of these models gets a semantics by mapping it from the language to a well-known semantic domain [1, 7]. This principle is well understood in the field of programming languages, where each syntactic construct has a well defined meaning that describes its effects in terms of operational or denotational semantics.

Although standardization bodies have not yet been able to define a commonly accepted, formal semantics, e.g., for the UML as yet, we here assume such a semantic definition would be given. See [7] for a deeper discussion on semantic issues. To understand the meaning of composition, it is evident that the meaning / semantics of the involved models needs to be understood.

### 2.1 Semantic Domain and Mapping

Given a language $M$ of models, the meaning of each element is usually given by explaining it in a well-known domain $D$, the semantic domain. This semantic domain describes which artifacts and concepts exist and must be well understood by both the language designer and the language users [7]. This principle is rather general, even so the details of the semantic domain as well as the form of representation vary. E.g., denotational as well as operational semantics can be subsumed under this form of approach using an abstract set of models resp. an abstract machine as semantic domain.

Examples for a semantic domain are the System Model [8], Abstract State Machines [9], or pure mathematics [2].

Given the modeling language $M$ and the semantic domain $D$ each model $m \in M$ must be mapped to $D$. As explained earlier, it is important to define the meaning (semantics) of models explicitly. So an explicit formal definition of the mapping is a function from $M$ to $D$:

$$sm: M \rightarrow D \qquad (1)$$

Benefits of a formal mapping function are that we are able to reason about the mapping and thus, about the language and the instances itself.

### 2.2 Set-Valued Semantics

A general problem of the semantics definition of a model is that models should be useable in early phases of development. In early phases models are usually underspecified and somewhat abstract. Therefore, there is usually not a single system that realizes a model, but a larger set of realizations. Thus, the mapping of an underspecified diagram to program code or any other deterministic realization would result in either incomplete code or code that incorporates decisions not present in the model. These decisions done by the translation algorithm, however, are critical for the model understanding, as they may not intend the developers view. Currently many tools help themselves, by disallowing ambiguity and thus preventing underspecification. A mapping to code, therefore, for principal reasons cannot serve as a semantics definition. To adequately handle underspecification the semantics of languages like Spectrum [10] or Z [11] is described as a set of systems having the given properties instead of a single system [12]. Such specification oriented set-valued semantics allow us to describe and understand important properties of modeling languages. Thus we use set-valued semantics as a basis for further investigation into a model composition theory.

The basic idea is to map any model $m \in M$ to all systems which obey the constraints that the model imposes. Denoting the set of all systems with $S$ the semantic domain is then the power set $D = \wp(S)$ and each instance $m \in M$ will be mapped by $sm$ to the largest set of systems which fulfill the constraints.

$$sm: M \to \wp(S) \qquad (2)$$

We do not need to further investigate into the details of $S$, but understand that it captures the relevant properties of a system. These are usually structural properties (objects, their values and linkage) as well as behavioral and interaction properties (traces of interactions, etc.).

As an illustrative example for set-valued semantics covering underspecification consider a simple class diagram with one class "Person" having a String attribute "name". What do we know about the system described?

1. There is a class "Person"
2. All instances of the class "Person" and all instances of subclasses have an attribute "name" whose type is "String"
3. No more information can be inferred.

The real semantics of this model must be given as the set of all systems obeying 1 and 2. Usually these systems have other classes and possibly the class "Person" contains more attributes than "name", but in our set-valued semantics those systems still fulfill the constraints defined by the model. Furthermore, it is not given that there will ever be an instance of class Person at all. Instead the class Person may also be abstract.

This approach is called a "loose semantics" [10] and is very helpful in capturing underspecification. Today many developers and especially tools assume some kind of "completeness" of their models, which is quite conflicting with the possibility to compose models.

Set-valued semantics allows to state some important properties with respect to the semantic mapping $sm$:

- A model $m \in M$ is **consistent** exactly if $sm(m) \neq \emptyset$, which means that there is at least one system that obeys the instance's properties. Otherwise, there are some contradicting constraints in the model $m$ itself.
- A model $m \in M$ does not contain information if $sm(m) = S$. Then any system can serve as an implementation.
- A model $m_2$ **refines** another model $m_1$ exactly if $sm(m_2) \subseteq sm(m_1)$. So, if we add more data to the model $m_2$, it further constraints the resulting set of systems, which therefore will become smaller.

The loose approach has an interesting aspect: the more we know, thus the more information is present in a model, the fewer implementations are possible. This is why $m_2$ has more information and thus refines $m_1$ exactly if $sm(m_2) \subseteq sm(m_1)$.

It is noteworthy that the "loose semantics" approach we use is loose on the behavioral as well as on the structural level. For existing behavioral elements, such as methods, their behavior may vary and additional structural elements (such as attributes, classes etc.) are possible.

Besides set-valued semantics for some forms of models and especially for executable languages an "initial" or a "minimal" semantics can be given. These forms of semantics correspond to the idea that there is a unique realization in the set mentioned above with minimalistic properties. Informally spoken, such a unique element can be characterized by assumptions like "everything explicitly defined is present, but nothing more". Class diagrams, e.g., lead to a canonical implementation through code generation and deterministic, completely defined state machines do have one single execution. Having both, a set-valued semantics for the specification of a system and an initial semantics, e.g., for test purposes or executable models, seems to be appropriate. For specification purposes, we concentrate on the set valued semantics.

## 3  An Algebraic View on Model Composition

When models are developed and composed, the developers as well as the tools always deal with their syntactic representation. But doing so, developers want to compose the meaning underlying these models. Thus, one goal of our algebraic theory is to clarify the relationships between composition on the syntactic and on the semantic level. Beyond that, another interesting issue consists in the question which basic requirements for a composition operator on the one hand and for composition tools on the other exist.

### 3.1  Model Composition

Model composition in its simplest form refers to the mechanism of combining two models into a new one. Without further information or requirements the definition of model composition is quite abstract. Denoting the universe of models with *M* we get the following definition of model composition operators:

**Definition 1: Model composition operator.**
A model composition operator $\otimes$ is a function with two models as input, which produces a composed model as output: $\otimes: \boldsymbol{M} \times \boldsymbol{M} \rightarrow \boldsymbol{M}$.

Given the semantics of models, we can infer properties of the semantics of a composition operator $\otimes$ by relating the semantics on its source and resulting model.

**Definition 2: Property preserving (PP) composition operator.**
A composition operator $\otimes: \boldsymbol{M} \times \boldsymbol{M} \rightarrow \boldsymbol{M}$ is property preserving on the left argument, if for any $\boldsymbol{m_1}, \boldsymbol{m_2} \in \boldsymbol{M}$ it holds: $sm(\boldsymbol{m_1} \otimes \boldsymbol{m_2}) \subseteq sm(\boldsymbol{m_1})$. Analogously, it is property preserving on the right argument, iff $sm(m_1 \otimes m_2) \subseteq sm(m_2)$ and **property preserving** (PP) if both properties hold.

The simple example shown in Figure 1 serves as basis for further explanations.

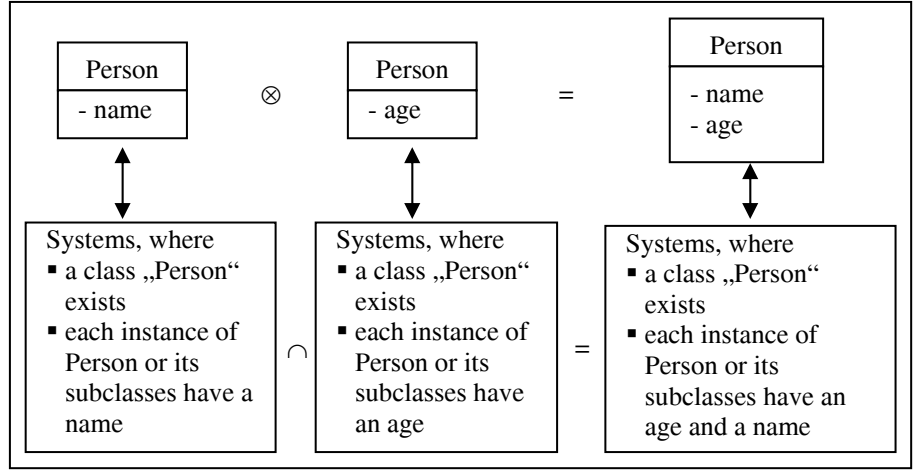

Figure 1: Example for composition on models and semantics

Property preservation is important for a composition operator, as it ensures that no information and thus, no design decisions that were present in a source model are lost in the composition. We can infer that property preservation is equivalent to:

$$\forall\ m_1, m_2 \in M:\ sm(m_1 \otimes m_2) \subseteq sm(m_1) \cap sm(m_2) \qquad (3)$$

Please note that this need not be equality, as the composition operator may be allowed to add further information that was not present in any of the models before. This can be useful, especially if there are decisions on unifications to make that are not unique. E.g., unnamed associations between the same classes can be identified, but need not.

**Definition 3: Fully property preserving (FPP) composition operator.**
A composition operator $\otimes: M \times M \rightarrow M$ is **fully property preserving**, iff

$$\forall\ m_1, m_2 \in M:\ sm(m_1 \otimes m_2) = sm(m_1) \cap sm(m_2) \qquad (4)$$

The most important consequence of FPP is that it allows us to separately analyze and understand the source models and their properties individually and to trace properties (as well as errors) of the composed model back to the input models. Furthermore, with a PP composition a model developer can be sure that the requirements defined in his models are preserved in the implementation. And third, a PP operator makes model composition understandable: changes in one input model have an impact on the composed model within a localized, clearly identifiable area, but do not affect properties defined in the other models.

A FPP composition operator neither adds nor forgets information. Unfortunately, we will have to live with the situation, that there are modeling languages, where there is no composed model that exhibits the desired properties. E.g., composing flat automata is not necessarily fully property preserving (depends on the assumed communication between these automata). In this case, emerging properties of the composition cannot necessarily be traced back to the original, but may result from the composition operator itself, which in fact is a composition and an additional refinement. However, adding wrong information through a composition operator may lead to an inconsistent result ($sm(m_1 \otimes m_2) = \varnothing$) even though the models originally where not inconsistent with each other ($sm(m_1) \cap sm(m_2) \neq \varnothing$). We therefore demand that composition preserves consistency:

**Definition 4: Consistency preserving (CP) composition operator.**
A composition operator $\otimes: M \times M \rightarrow M$ is **consistency preserving (CP)**, iff

$$\forall\ m_1, m_2 \in M:\ sm(m_1) \cap sm(m_2) \neq \varnothing\ \Rightarrow\ sm(m_1 \otimes m_2) \neq \varnothing \quad (5)$$

**Corollary:** A FPP composition operator is consistency preserving.
*Proof*: by definition.

In general as well as in the remainder of this paper we assume model composition to be property preserving as well as consistency preserving (but not in all cases fully property preserving).

### 3.2 A Generalization for Semantic Composition Operators

We have explained the desired properties of a composition operator using set-valued semantics. This technique can be generalized, assuming there is a composition operator $\oplus$ available on the semantic domain. Intersection $\cap$ as used above is such an operator.

**Definition 5: General Semantic Composition Operator.**
The semantic composition operator $\oplus$ is a function with two sets of systems as input which produces a set of systems as output: $\oplus: \boldsymbol{D} \times \boldsymbol{D} \rightarrow \boldsymbol{D}$.

Given these operators on both levels, the semantic composition operator $\oplus$ can be understood as semantics of the syntactic operator $\otimes$ if the diagram in Figure 2 commutes.

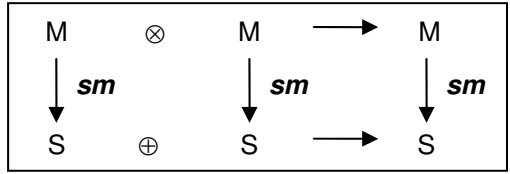

Figure 2: Relationship between composition operators

We say the diagram commutes iff

$$\forall\, m_1, m_2 \in M:\ sm(m_1 \otimes m_2) = sm(m_1) \oplus sm(m_2) \tag{6}$$

A commuting diagram corresponds to a fully property preserving composition as defined above and exhibits the same advantages as discussed above. We therefore impose the requirement that the diagram in Figure 2 should always commute. If not, at least the relaxed version must be considered:

$$\forall\, m_1, m_2 \in M:\ sm(m_1 \otimes m_2) \subseteq sm(m_1) \oplus sm(m_2) \tag{7}$$

Therefore, the syntactic operator $\otimes$ reflects the semantic composition $\oplus$ and an additional refinement. However, in the following we use intersection as semantic composition only.

### 3.3 Syntax-Based Properties of Composition

Examining properties of the syntactic composition $\otimes$, we find that there may be absorbing or neutral elements. In a first attempt, we may call a model $m \in M$ right-neutral, iff

$$\forall\, m_1 \in M:\ m_1 \otimes m = m_1 \tag{8}$$

A model $m \in M$ is called right-absorbing, iff

$$\forall\, m_1 \in M:\ m_1 \otimes m = m \tag{9}$$

Left-neutral and left-absorbing is defined analogously and **neutral** respectively **absorbing** is the combination of both sides. Furthermore, we might call a composition operator $\otimes$ **commutative** iff

$$\forall\, m_1, m_2 \in M:\ m_1 \otimes m_2 = m_2 \otimes m_1 \tag{10}$$

and **associative** iff

$$\forall\, m_1, m_2, m_3 \in M:\ (m_1 \otimes m_2) \otimes m_3 = m_1 \otimes (m_2 \otimes m_3) \tag{11}$$

Of course, if the composition operator is commutative, left and right-neutrality as well as properties to be left-/right-absorbing will coincide.

There may be many models that are absorbing or neutral. But, due to unlucky context conditions there may also be none at all. For class diagram composition, a neutral element could be the empty class diagram, which is not allowed in UML 2.1.

This formalization above would allow us to identify an algebra of composition on the syntactic level. However, when looking at the properties, we easily can see that this algebra is too restrictive to be of direct use. In fact models have a concrete syntax and the positions of white spaces or the graphical elements usually change, when models are composed or somehow otherwise modified. Furthermore, the order of presenting elements usually does not affect the semantics, but the layout of the composed result. An example in Figure 3 shows a possible key problem.

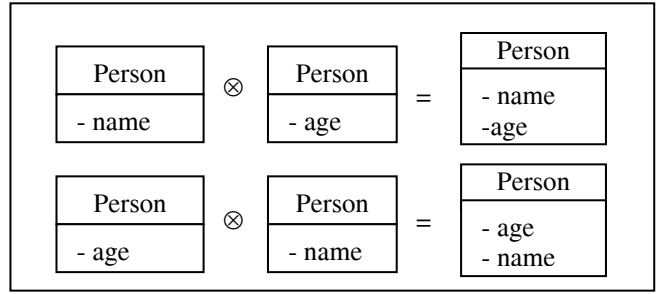

Figure 3: Example for non-commutative model composition (on syntactic level)

This example leads us to two observations. First, the result syntactically depends on the order of the input models and thus, composition is often not commutative. Second, the result does not depend semantically on the input order, since the outputs are "semantically equal", which means that they are mapped by *sm* to the same set of systems. Therefore, we do generalize from a purely model (syntax)-based concept of composition to a semantic-based version.

### 3.4 Semantic-Based Composition Properties

Instead of defining associativity, etc. on the concrete syntax of models, we abstract away from irrelevant syntactic sugar and concentrate on the semantic properties of a model. Therefore, we develop the following definitions:

**Definition 6: Algebraic Properties of Composition.**
A model $m \in M$ is called right-**neutral** vs. composition $\otimes$, iff

$$\forall\, m_1 \in M: sm(m_1 \otimes m) = sm(m_1) \tag{12}$$

Model $m \in M$ is called right-**absorbing** vs. composition $\otimes$, iff

$$\forall\, m_1 \in M: sm(m_1 \otimes m) = sm(m) \tag{13}$$

A model $m \in M$ is called right-**idempotent** vs. composition $\otimes$, iff

$$\forall\, m_1 \in M: sm((m_1 \otimes m) \otimes m) = sm(m_1 \otimes m) \tag{14}$$

Being left-neutral, -absorbing and –idempotent is defined in an analogous way.
If a model is neutral (absorbing/idempotent) from both sides, it is called neutral (absorbing/idempotent).

We call a composition operator $\otimes$ **commutative** vs. its semantics *sm* iff

$$\forall\ m_1, m_2 \in M: sm(m_1 \otimes m_2) = sm(m_2 \otimes m_1) \tag{15}$$

and **associative** vs. its semantics *sm* iff

$$\forall\ m_1, m_2, m_3 \in M: sm((m_1 \otimes m_2) \otimes m_3) = sm(m_1 \otimes (m_2 \otimes m_3)) \tag{16}$$

This formalization allows us to define an algebra with composition etc. based on semantic properties. Looking at these properties from a different angle, we can identify an equivalence relation $\cong$ on models based on the semantic mapping interpreted as homomorphism.

### 3.5 Properties of the Semantic Mapping

Let in this section $\otimes$ be a FPP composition operator. We know that $(\wp(S), \cap, S, \varnothing)$ defines a lattice, where intersection is both commutative and associative. Together with the semantic mapping *sm* we can translate the lattice properties to the language of models:

**Theorem 1:**
If a model composition $\otimes$ is fully property preserving, then $(M, \otimes)$ also defines a commutative, associative structure with respect to *sm* and $\otimes$ is idempotent for all models.

*Proof*: By definition of FPP we derive
Assoc.: $sm((m_1 \otimes m_2) \otimes m_3) = sm(m_1) \cap sm(m_2) \cap sm(m_3) = sm(m_1 \otimes (m_2 \otimes m_3))$,
Comm.: $sm(m_1 \otimes m_2) = sm(m_1) \cap sm(m_2) = sm(m_2 \otimes m_1)$, and
Idempot.: $sm(m_1 \otimes m_1) = sm(m_1) \cap sm(m_1) = sm(m_1)$. $\square$

Respecting the semantic equivalence of two models is an important property for a composition operator, because then the concrete representative is irrelevant and layout or other minor rearrangements of the model do not affect the composition result. We therefore introduce the algebra of equivalence classes on models induced by the semantic mapping:

**Definition 7: Equivalence Classes of Models**
The semantic mapping *sm* defines an equivalence relation on models as follows:

$$m_1 \cong m_2 \iff sm(m_1) = sm(m_2) \tag{17}$$

The set of semantically equivalent models is denoted by

$$[m_1] = \{\ m_2\ |\ m_1 \cong m_2\ \} \tag{18}$$

We denote the set of equivalence classes over $M$ by $[M]$. The composition operation can be extended to equivalence classes as follows:

**Definition 8: Composition on Model Classes**
Composition is extended to model classes by:

$$[m_1] \otimes [m_2] = \{ m_a \otimes m_b \mid m_a \in [m_1] \wedge m_b \in [m_2] \} \quad (19)$$

**Theorem 2: [.] is a congruence for FPPs**
If a model composition $\otimes$ is fully property preserving, then $([M], \otimes)$ also defines a commutative, associative structure with respect to *sm*, all models are idempotent, and:

$$[m_1] \otimes [m_2] = [m_1 \otimes m_2] \quad (20)$$

*Proof*: Follows from FPP and the definition of the equivalence classes. □

We now have a quotient algebra $([M], \otimes)$ with a number of desired properties for a syntactic composition operator:
1. Composition is fully property preserving, such that each property of the composed model can be traced back to one of the input models or both.
2. Composition is consistent with the semantics, such that it is irrelevant, which concrete representative was chosen. Thus the composition is well defined with respect to the quotient algebra.
3. Composition is commutative and associative, such that the order of composition is irrelevant.

As already discussed, unfortunately a number of composition operators will exist that do not fit this ideal scheme for a variety of reasons. E.g., it may rather often be the case that an operator is PP and CP, but not FPP. In this case, it may happen that even if the operator is commutative and associative on models, the equivalence on models is not a congruence vs. composition.

A model composition operator which depends on the order of the input or concrete representations of the model would be difficult to manage. E.g., the input order has to be saved somewhere to guarantee the equality of the results.

From theoretical computer science, we know that composition operators need to conform with semantics as much as possible. This may be achieved through a number of mechanisms. On the one hand the composition operator may be adjusted accordingly. Second, the semantic domain or the semantic mapping may be redefined, such that they go conform with composition and third, the modeling language itself may be adapted.

### 3.6 Summary

In the last sections we introduced some basic properties model composition operators may have such as PP, FPP, or CP. Following we give a short overview of the definitions which allow to categorize a given composition operator.

| Property | Requirement | Dependencies |
|---|---|---|
| Property Preserving on the left (PP$_l$) | $sm(m_1 \otimes m_2) \subseteq sm(m_1)$ | |
| Property Preserving on the | $sm(m_1 \otimes m_2) \subseteq sm(m_2)$ | |
| right (PP$_r$) | | |
| Property Preserving (PP) | $sm(m_1 \otimes m_2) \subseteq sm(m_1) \cap sm(m_1)$ | PP$_l \wedge$ PP$_r \Leftrightarrow$ PP |
| Fully Property Preserving | $sm(m_1 \otimes m_2) = sm(m_1) \cap sm(m_2)$ | FPP $\Rightarrow$ PP |
| Consistency Preserving | $\forall m_1, m_2 \in M: sm(m_1) \cap sm(m_2) \neq \emptyset \Rightarrow sm(m_1 \otimes m_2) \neq \emptyset$ | FPP $\Rightarrow$ CP |
| Commutative (Com) | $\forall m_1, m_2 \in M: m_1 \otimes m_2 = m_2 \otimes m_1$ | |
| Associative (Ass) | $\forall m_1, m_2, m_3 \in M:$ $(m_1 \otimes m_2) \otimes m_3 = m_1 \otimes (m_2 \otimes m_3)$ | |
| Commutative vs. Semantics (Com$_{sm}$) | $\forall m_1, m_2 \in M:$ $sm(m_1 \otimes m_2) = sm(m_2 \otimes m_1)$ | Com $\Rightarrow$ Com$_{sm}$ |
| Associative vs. Semanics (Ass$_{sm}$) | $\forall m_1, m_2, m_3 \in M: sm((m_1 \otimes m_2) \otimes m_3) = sm(m_1 \otimes (m_2 \otimes m_3))$ | Ass $\Rightarrow$ Ass$_{sm}$ |

**Table 1: Overview of Composition properties**

Furthermore, we defined special elements with respect to composition. Table 2 gives a short overview.

| Property of Element $m$ | Requirement | Dependencies |
|---|---|---|
| Right-neutral (Rn) | $\forall m_1 \in M: m_1 \otimes m = m_1$ | |
| Left-neutral (Ln) | $\forall m_1 \in M: m \otimes m_1 = m_1$ | |
| Neutral (N) | $\forall m_1 \in M: m_1 \otimes m = m \otimes m_1 = m_1$ | Rn $\wedge$ Ln $\Leftrightarrow$ N |
| Right-absorbing (Ra) | $\forall m_1 \in M: m_1 \otimes m = m$ | |
| Left-absorbing (La) | $\forall m_1 \in M: m \otimes m_1 = m$ | |
| Absorbing (A) | $\forall m_1 \in M:$ $m_1 \otimes m = m \otimes m_1 = m$ | Ra $\wedge$ La $\Leftrightarrow$ A |
| Right-Idempotent (Ri) | $\forall m_1 \in M:$ $(m_1 \otimes m) \otimes m = m_1 \otimes m$ | |
| Left-Idempotent (Li) | $\forall m_1 \in M:$ $m \otimes (m \otimes m_1) = m_1 \otimes m$ | |
| Idempotent (I) | $\forall m_1 \in M: m \otimes (m \otimes m_1) = (m_1 \otimes m) \otimes m = m_1 \otimes m$ | Ri $\wedge$ Li $\Leftrightarrow$ I |
| Right-neutral vs. Composition (Rn$_{comp}$) | $\forall m_1 \in M:$ $sm(m_1 \otimes m) = sm(m_1)$ | Rn $\Rightarrow$ Rn$_{comp}$ |
| Left-neutral vs. Composition (Ln$_{comp}$) | $\forall m_1 \in M:$ $sm(m \otimes m_1) = sm(m_1)$ | Ln $\Rightarrow$ Ln$_{comp}$ |
| Neutral vs. Composition (N$_{comp}$) | $\forall m_1 \in M:$ $sm(m_1 \otimes m) = sm(m \otimes m_1) = sm(m_1)$ | Rn$_{comp} \wedge$ Ln$_{comp} \Leftrightarrow$ N$_{comp}$ N $\Rightarrow$ N$_{comp}$ |
| Right-absorbing vs. Composition (Ra$_{comp}$) | $\forall m_1 \in M:$ $sm(m_1 \otimes m) = sm(m)$ | Ra $\Rightarrow$ Ra$_{comp}$ |
| Left-absorbing vs. Composition (La$_{comp}$) | $\forall m_1 \in M:$ $sm(m \otimes m_1) = sm(m)$ | La $\Rightarrow$ La$_{comp}$ |

| Absorbing vs. Composition ($A_{comp}$) | $\forall m_1 \in M:$ $sm(m_1 \otimes m) = sm(m \otimes m_1) = sm(m)$ | $Ra_{comp} \wedge La_{comp} \Leftrightarrow A_{comp}$ $A \Rightarrow A_{comp}$ |
|---|---|---|
| Right-Idempotent vs. Composition ($Ri_{comp}$) | $\forall m_1 \in M:$ $sm((m_1 \otimes m) \otimes m) = sm(m_1 \otimes m)$ | $Ri \Rightarrow Ri_{comp}$ |
| Left-Idempotent vs. Composition ($Li_{comp}$) | $\forall m_1 \in M:$ $sm(m \otimes (m \otimes m_1)) = sm(m_1 \otimes m)$ | $Li \Rightarrow Li_{comp}$ |
| Idempotent vs. Composition ($I_{comp}$) | $\forall m_1 \in M:$ $sm(m \otimes (m \otimes m_1)) = sm((m_1 \otimes m) \otimes m) = sm(m_1 \otimes m)$ | $Ri_{comp} \wedge Li_{comp} \Leftrightarrow I_{comp}$ $I \Rightarrow I_{comp}$ |

**Table 2: Special elements of Composition**

## 4 Related Work

Much work on specification with respect to model composition has been done in the formal methods community. Based on [13] the notion of fully abstract composition was transferred to a number of formal languages for behavioral specification. Our approach is very much in the spirit of this work, but tries to identify interesting sub-properties for model composition as well.

Model composition is also a widespread research issue in the world of UML. There are several works which concentrate on different kinds of UML-like diagrams, as class diagrams [14] or state charts [15]. Most of these works do not discuss composition or model management operators from a foundational, algebraic point of view and thus, have different objectives.

In [4] three model composition tools, namely the Atlas Model Weaver, the Epsilon Merging Language, and the Glue Generator Tool which were developed in the Modelware project [18] are introduced and discussed in detail. Furthermore, it derives some common definitions from these discussions and clarifies some basic requirements for model composition tools and frameworks. However, our work concentrates on the semantic properties of model composition, whereas [4] addresses mainly syntactic properties and their implementation in tools.

A generic semantics of the merge operator was presented in the MOMENT project [19]. It describes three steps of model merging: finding semantic equivalences, conflict resolution, and copying non-duplicated elements. In contrast to our work it concentrates on expressing semantic equalities by means of a metamodel whereas we discuss the semantic background of model composition.

A more theoretical view on different model management operators is presented in [16]. It introduces algebraic properties of model merging such as commutativity, associativity, and idempotency. The theoretical results are illustrated by two examples, merging entity relationship models and state machines, respectively. In opposition to our work the algebra of model composition is not discussed in detail.

Instead the concentration lies on a general overview of model management operators and their relationships.

An algebra of merging incomplete and inconsistent graph-based views is discussed in [17]. Category theory and colimits serve as theoretical basis to express the relationships between different diagrams in opposition to our viewpoint of algebras. Furthermore, the basic intention of [17] consists in the identification of equal elements in different views whereas our work concentrates on the algebraic properties of model composition.

## 5   Conclusion

In this paper we gave a first contribution to shed light into the question how model composition operators interact with the semantics of models and what properties composition operators should have. For this purpose, we have abstractly described how semantics is defined. We then introduced an algebra of model composition that describes the formal relationship between the models, equivalence classes of semantically equivalent models, model composition and semantics. From this setting some results could be derived. The most important are that model composition should be a congruence induced by the semantic definition and a composition should be a commutative and associative operator with respect to the semantics.

These theoretical results lead to practical consequences for the design of model composition operators, modeling languages and semantic domains. Any composition operator should obey the properties implied by the algebra in order to allow a modular model-based development of software systems with independent compilation/transformation of models to other representations and levels of abstraction.

This paper is concerned with the model composition operator and its implications. It can be seen as a foundation for further investigations on model management operations. However, there are a number of extensions to deal with: How to deal with a diff operator to reverse composition? How does code and test-case generation interact with composition and semantics? Are there impacts for the form of meta-modeling widely used today? What are properties of an unsymmetric composition like aspect weaving? How do UML's semantic variation points interact with composition? Will refinement preserving composition be useful and feasible? Will there be compositional refactorings? Many of these questions need to be solved for a foundational theory of model composition.

**Acknowledgement:** The work presented in this paper is undertaken as a part of the MODELPLEX project. MODELPLEX is a project co-funded by the European Commission under the "Information Society Technologies" Sixth Framework Programme (2002-2006). Information included in this document reflects only the authors' views. The European Community is not liable for any use that may be made of the information contained herein.